\documentclass[reprint, amsmath, amssymb, superscriptaddress, prb]{revtex4-2}
\usepackage{graphicx}
\usepackage{hyperref}
\usepackage[dvipsnames]{xcolor}
\usepackage{bbm}

\begin{document}

\title{Compact inductor-capacitor resonators at sub-gigahertz frequencies}
\author{Qi-Ming Chen}
\email{qiming.chen@aalto.fi}
\affiliation{QCD Labs, QTF Centre of Excellence, Department of Applied Physics, Aalto University, FI-00076 Aalto, Finland}

\author{Priyank Singh}
\author{Rostislav Duda}
\author{Giacomo Catto}
\author{Aarne Ker{\"a}nen}
\author{Arman Alizadeh}
\author{Timm M{\"o}rstedt}
\author{Aashish Sah}
\author{Andr{\'a}s Gunyh{\'o}}
\affiliation{QCD Labs, QTF Centre of Excellence, Department of Applied Physics, Aalto University, FI-00076 Aalto, Finland}

\author{Wei Liu}
\affiliation{QCD Labs, QTF Centre of Excellence, Department of Applied Physics, Aalto University, FI-00076 Aalto, Finland}
\affiliation{IQM Quantum Computers, Espoo 02150, Finland}


\author{Mikko M{\"o}tt{\"o}nen}
\affiliation{QCD Labs, QTF Centre of Excellence, Department of Applied Physics, Aalto University, FI-00076 Aalto, Finland}
\affiliation{VTT Technical Research Centre of Finland Ltd. $\&$ QTF Centre of Excellence, P.O. Box 1000, 02044 VTT, Finland}

\date{\today}
\begin{abstract}
Compact inductor-capacitor (LC) resonators, in contrast to coplanar waveguide (CPW) resonators, have a simple lumped-element circuit representation but usually call for sophisticated finite-element method (FEM) simulations for an accurate modelling. Here we present an simple analytical model for a family of coplanar LC resonators where the electrical properties are directly obtained from the circuit geometry with a satisfying accuracy. Our experimental results on $10$ high-internal-quality-factor resonators ($Q_{\rm i}\gtrsim 2\times 10^{5}$), with frequency ranging roughly from $300\,{\rm MHz}$ to $1\,{\rm GHz}$, show an excellent consistency with both the derived analytical model and detailed FEM simulations. These results showcase the ability to design sub-gigahertz resonators with less than $2\%$ deviation in the resonance frequency, which has immediate applications, for example, in the implementation of ultrasensitive cryogenic detectors. The achieved compact resonator size of the order of a square millimeter indicates a feasible way to integrate hundreds of microwave resonators on a single chip for realizing photonic lattices.
\end{abstract}

\maketitle

\section{Introduction}
Superconducting quantum circuits (SQC) provide a versatile platform of quantum engineering that has led to groundbreaking results in quantum-microwave communication\,\cite{Axline2018, Kurpiers2018, Pogorzalek2019, Fedorov2021}, computation\,\cite{DiCarlo2009, Lucero2012, Zheng2017, Chen2020, Harrigan2021}, simulation\,\cite{Underwood2012, AbdumalikovJr2013, Roushan2016, Roushan2017, Kollar2019, Ma2019, Xu2020, Guo2020}, and sensing\,\cite{Barzanjeh2020, Bienfait2017, Wang2021, Govenius2016, Kokkoniemi2019, Kokkoniemi2020, Gasparinetti2015, Karimi2020}. Besides the celebrated Josephson effects, the rapid development of this field may be largely attributed to the wide application of coplanar waveguides (CPWs)\,\cite{Wallraff2004, Majer2007}, especially CPW resonators, which are ubiquitous throughout the SQC technology. 

\begin{figure}[t]
  \centering
  \includegraphics[width=\columnwidth]{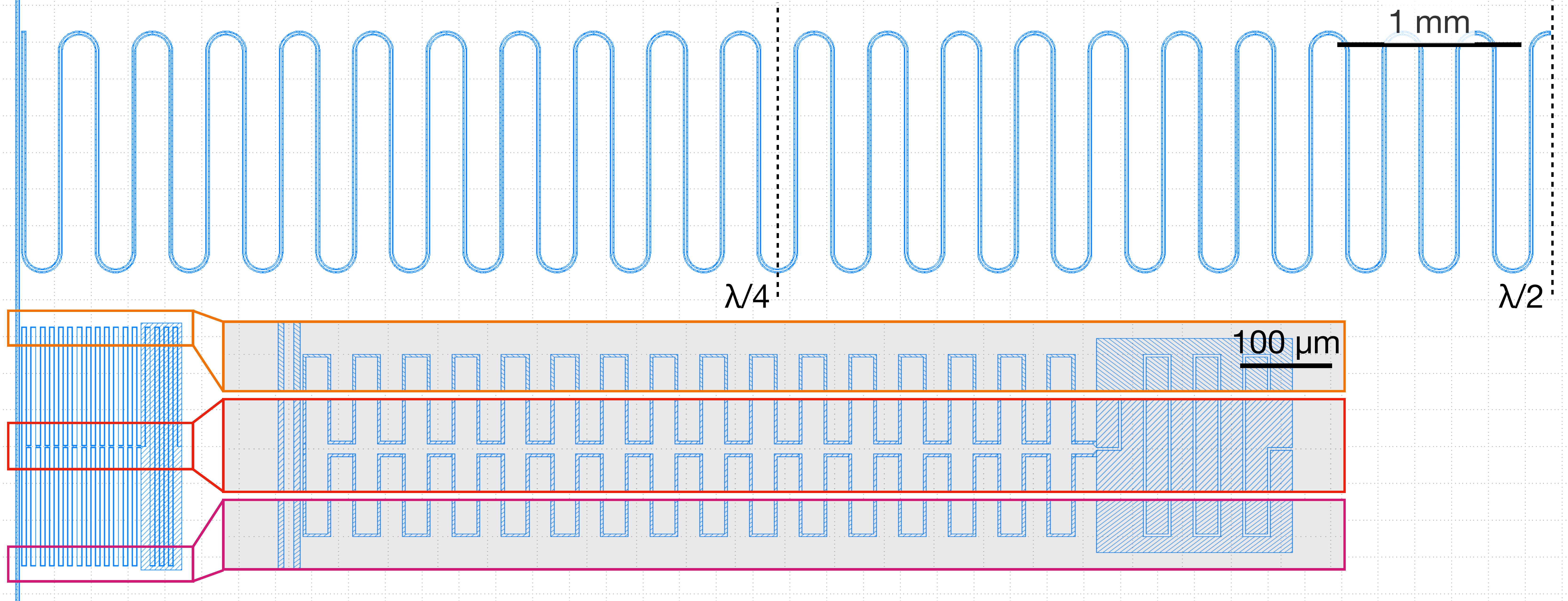}
  \caption{Example of a CPW resonator (top) and a compact inductor-capacitor (LC) resonator (bottom) of an identical lowest resonance frequency of $1\,{\rm GHz}$. White color indicates superconducting thin-film metal viewed from the top and blue color denotes regions where the metal is etched away, exposing the substrate below the metal. Both resonators are coupled to a transmission line on the left of the structure. The width and gap (gap and width) of the capacitor (inductor) element in the LC resonator are chosen as $22\,{\rm \mu m}$ and $3\,{\rm \mu m}$, respectively, for the ease of fabrication.  The center-conductor width and gap of the CPW resonator are $10\,{\rm \mu m}$ and $6\,{\rm \mu m}$, respectively, for achieving a $50$-${\rm \Omega}$ characteristic impedance, yielding the total electrical length of $59\,{\rm mm}$.}
  \label{fig_schematic}
\end{figure}

A CPW resonator is essentially a transmission line with short- or open-circuit boundary conditions. They can be modelled analytically, and are flexible in design and relatively simple to fabricate\,\cite{Frunzio2004, Goeppl2008}. In SQC, the CPW resonators have a physical size comparable to or larger than the wavelength of the microwave field, $\lambda \approx 20\,{\rm mm}$ at $5\,{\rm GHz}$, making them particularly useful for circuits resonating in the several of gigahertz regime. However, at sub-gigahertz frequencies, the relative permittivity of typical low-loss substrate materials, such as Si with $\epsilon_{\rm r} \approx 11.9$, implies a CPW resonator to have a length of roughly $100\,{\rm mm}$. Such a large structure shown in Fig.\,\ref{fig_schematic} (top) leads to severe limitations on the number of resonators that can be fit on a single chip. Winding the waveguide into a spiral shape may be a convenient solution in certain cases\,\cite{Zhuravel2012, Maleeva2014, Maleeva2015, Partanen2016, Rolland2019, Yan2021}, but it also raises new challenges such as impedance matching and grounding over such a long distance\,\cite{Wenner2011, Lankwarden2012, Abuwasib2013, Chen2014}. Other issues, such as coupling to the parasitic modes of the sample holder\,\cite{Fischer2021}, should also be treated with extra care to design a huge structure.

An alternative, but less explored way for making superconducting resonators, is to use the lumped-element circuits such as the interdigital capacitors (IDCs) and meander-line inductors (MLIs). These circuits have a compact physical size being much smaller than the wavelength of the microwave field\,\cite{Lindstroem2009, Leduc2010, Khalil2011, Geerlings2012, McKay2015, Weichselbaumer2019, Peruzzo2020}, as shown in Fig.\,\ref{fig_schematic} (bottom). The lumped-element inductor-capacitor (LC) resonators can be well described by a simple lumped-element parallel or series circuit consisting of an inductor, $L$, and a capacitor, $C$. In addition to the compactness, the internal quality factor of a lumped-element resonator is typically high \,\cite{Lindstroem2009}. The characteristic impedance of a lumped-element resonator is also easy to adjust in a large range, whereas that of a CPW resonator is normally upper bounded by approximately $377\,{\rm \Omega}$ \cite{Peruzzo2020}. However, estimating the precise values of $L$ and $C$ in a lumped-element resonator is generally challenging without using the finite-element method (FEM), which hinders their applications in SQC. The FEM simulations are especially resource-intensive for the lumped-element circuits, where the fine structures should be taken good care of with smal-enough mesh size. Furthermore, numerical results provide little intuition on how to adjust the geometry to obtain the desired values of $L$ and $C$, respectively. Thus, there is a great need in lumped-element-resonator design to estimate the electrical properties from the circuit geometry using convenient analytical equations, as explicitly called for in the recent studies of SQC\,\cite{McKay2015, Weichselbaumer2019}.

Although the values of $L$ and $C$ should be fully determined from the circuit layout, an accurate estimation of them is available only in simple cases. Fortunately, practical approximations for certain structures can already be found in the existing literature on microwave engineering\,\cite{Grover1946, Greenhouse1974, Yue2000, Hsu2004, Wei1977, Igreja2004, Bao2019}. It is therefore possible to model a compact LC resonator by synthesizing these early works in the specific context of SQC. In this study, we are particularly interested in the coplanar structures, such as MLIs and IDCs, as they are straightforwardly compatible with the thin-film technology used in SQC\,\cite{Grover1946, Greenhouse1974, Yue2000, Hsu2004, Wei1977, Igreja2004, Bao2019}. We first scrutinize the most useful design rules of MLIs and IDCs into compact analytical equations, and verify them in the typical parameter range of SQC via FEM. We then apply this knowledge to design sub-gigahertz LC resonators without FEM, with resonance frequencies ranging from $300\,{\rm MHz}$ to $1\,{\rm GHz}$ and impedances from $30\,{\rm \Omega}$ to $60\,{\rm \Omega}$. Subsequently, we implement the designs with one-step laser lithography and measure two samples at the cryogenic temperature of roughly $30\,{\rm mK}$. Our experimental results show an excellent agreement with both the analytical model and the FEM simulations. These results provide a systematic study of compact LC resonators in SQC, and hence fill the sub-gigahertz frequency gap in the design of superconducting microwave resonators. The compact size and the high internal quality factor of the realized LC resonators is expected to advance the development of ultrasensitive microwave bolometers and calorimeters\,\cite{Govenius2016, Kokkoniemi2019, Kokkoniemi2020, Gasparinetti2015, Karimi2020}. The achieved quare-millimeter footprint also indicates a feasible way to build a large microwave resonator array on a single chip for quantum information processing and quantum simulations\,\cite{Ozawa2019}. 

\section{Meander-line inductor (MLI)}
\begin{figure}[t]
  \centering
  \includegraphics[width=\columnwidth]{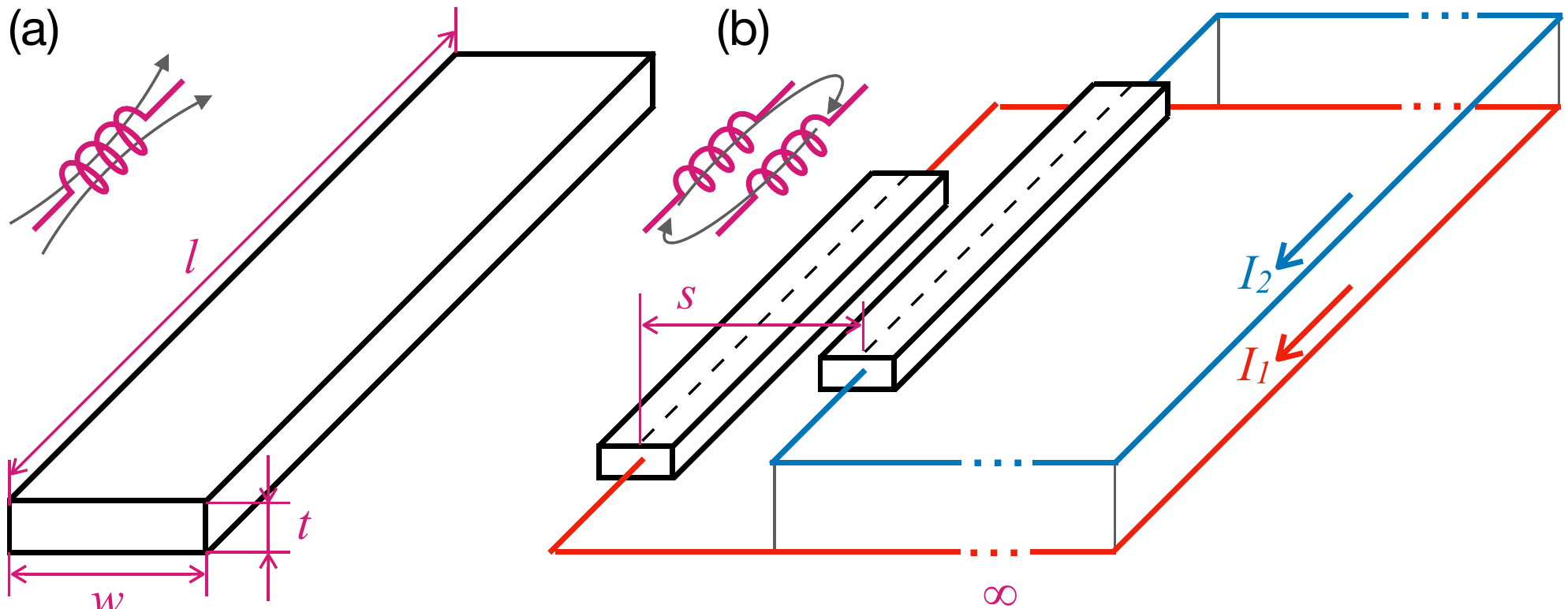}
  \caption{Illustration of self and mutual inductances for a MLI. (a) The self inductance of a single metal strip with width $w$, length $l$, and thickness $t$. (b) The mutual inductance between two parallel strips with a pitch distance of $s$. Here, the circuit can be visualized as a solenoid which is formed by the parallel wires and the ground at an infinitely far distance. Depending on the direction of the currents flowing through the two wires, $I_1=\pm I_2$, the mutual inductance can be either positive or negative.}
  \label{fig_inductance} 
\end{figure}

We consider a MLI which is made of a superconducting meandering wire. The current going through the wire creates a magnetic field that inhibits the change of the current itself, corresponding to a self inductance, $L_{\rm s}$ [Fig.\,\ref{fig_inductance}(a)]. Because of the meander-line geometry, the magnetic fields generated by local parallel lines may also contribute to each other and lead to a mutual inductance, $L_{\rm m}$ [Fig.\,\ref{fig_inductance}(b)]. We therefore describe the total inductance of the MLI element as
\begin{align}
	L = L_{\rm s} + L_{\rm m}. \label{eq:inductance}
\end{align}
The mutual inductance can be either positive or negative depending on the direction of current in the two parallel lines. The sign of $L_{\rm m}$ is determined by the so-called flux linkage, which is formally defined by an integral of the magnetic field over a Riemann surface bounded by the entire circuit. A simple visualization of the flux linkage is to imagine a returning current at an infinitely far distance, such that the two wires and the returning current forms a solenoid, as schematically shown in Fig.\,\ref{fig_inductance}(b). The rule of thumb is to take the positive (negative) sign when the current in two parallel lines under consideration are pointing in the same (opposite) direction\,\cite{Greenhouse1974}.

\subsection{Self inductance}
In this section, we illustrate the self inductance calculation with a simple geometry which leads to exact analytical solutions \cite{Grover1946}. Here, a current $I$ is flowing  along the $y$ axis in a cylindrical conductor of radius $R$ and length $l$. The magnetic field at an arbitrary position $\mathbf{r}$ in the $xy$-plane outside the conductor can be calculated with the Biot--Savart law 
\begin{align}
	\mathbf{B}(\mathbf{r}) &= 
	\frac{\mu_{0}I}{4\pi}\int_{0}^{l}\frac{\mathrm{d}\hat{\mathbf{y}}\times (\mathbf{r}-\mathbf{y})}{|\mathbf{r}-\mathbf{y}|^3} \nonumber\\
	&= \frac{\mu_{0}I}{4\pi}\left(\frac{l-y}{x\sqrt{x^2 + (l-y)^2}}
	+ \frac{y}{x\sqrt{x^2 + y^2}}\right)\hat{\mathbf{z}},
\end{align}
where $\hat{\mathbf{x}}$, $\hat{\mathbf{y}}$, and $\hat{\mathbf{z}}$ are the otrhonormal Cartesian unit vectors, and $\mu_{0}=4\pi\times 10^{-7}\,{\rm H/m}$ is the permeability of free space. The magnetic flux penetrating the $xy$ plane along the length of the conductor is thus
\begin{align}
	\Phi_{\rm ext} &= \int_{R}^{\infty}\int_{0}^{l}  B(\mathbf{r})\,\mathrm{d}x\,\mathrm{d}y \nonumber\\
	&= \frac{\mu_{0}I}{2\pi}\left[l\ln\left(\frac{l+\sqrt{l^2+R^2}}{R}\right) + R - \sqrt{l^2+R^2}\right] \nonumber\\
	&\approx \frac{\mu_{0}Il}{2\pi}\left[\ln\left(\frac{2l}{R}\right) - 1 \right],
\end{align}
where we have assumed $l \gg R$ in the last row for simplification. This flux gives rise to the external part of the self inductance
\begin{align}
	L_{\rm ext} \approx \frac{\mu_{0}l}{2\pi}\left[\ln\left(\frac{2l}{R}\right) - 1 \right].
\end{align}

Inside the wire and under the low-frequency condition, the current is uniformly distributed over the cross section with little contribution of the here-neglected skin effect. The current that flows within a round-shaped cross section or radius $x$ is thus $Ix^2/R^2$. The fact that $I$ flows in the $y$ direction indicates that the magnetic field must be parallel to the $xy$ plane, while the symmetry argument indicates that the field lines form circles centered at the $y$ axis. In this regard, we simply apply Amp{\`e}re's law and obtain 
\begin{align}
	\mathbf{B}(x) = \frac{\mu_{0}I x}{2\pi R^2}\hat{\mathbf{z}}.
\end{align} 
We can therefore calculate the so-called flux linkage as
\begin{align}
	\Phi_{\rm int} &= \int_{0}^{R}\int_{0}^{l}\frac{x^2}{R^2}\,B(x)\mathrm{d}x\,\mathrm{d}z 
	= \frac{\mu_{0}I l}{8\pi},
\end{align}
and therefore the internal part of the self inductance
\begin{align}
	L_{\rm int} = \frac{\mu_{0}l}{8\pi}.
\end{align}
Here, the factor $x^2/R^2$ is introduced in the integral because the magnetic field created by the enclosed current, $Ix^2/R^2$, does not occupy the entire cross section of the wire\,\cite{Grover1946}. It is this factor that distinguishes the concept of flux linkage from the physical magnetic flux through a simple surface. These two concepts are equivalent to each other by introducing an appropriately chosen Riemann surface. In a solenoid, for example, such a Riemann surface is bounded by the winding wire, leading to the flux generated by a single turn to thread the round surface many times, thus enhancing the flux linkage. Here, we may visualize the factor, $x^2/R^2$, as a fractional number of turns that the magnetic field penetrates the surface. 

In total, we have the self inductance of the wire as
\begin{align}\label{eq:inductance_cylinder}
	L_{\rm s} = L_{\rm int}+L_{\rm ext}
	\approx \frac{\mu_{0}l}{2\pi}\left[\ln\left(\frac{2l}{R}\right) 
	-\frac{3}{4}\right].
\end{align}
We note that the above equation does not take the penetration depth into consideration, since the size of the cross section is assumed to be much larger than the penetration depth for fabrication ease. The discussion of superconducting nanowires is beyond the interest of this example and may be found elsewhere\,\cite{Hopkins2005, Mooij2006, Samkharadze2016, Murphy2017, Hazard2019}.

We now express the total self inductance, $L_{\rm s}$, in a slightly different form as
\begin{align}
	L_{\rm s} = \frac{\mu_{0}l}{2\pi}\left[\ln\left(\frac{2l}{r_{\rm GMD}}\right) - 1 + \frac{r_{\rm AMD}}{l}\right], \label{eq:inductance_self_general}
\end{align}
where $r_{\rm GMD}=e^{-1/4}R$ and $r_{\rm AMD}=R$ are the geometric and algebraic mean distances between all the points inside the cylinder wire, respectively\,\cite{Grover1946, Greenhouse1974}. We note that Eq.\,\eqref{eq:inductance_cylinder} can be recovered in the limit $R \ll l$. Conveniently, Eq.~\eqref{eq:inductance_self_general} may be generalized to conductors with an arbitrary cross section. In SQC, we are particularly interested in coplanar circuits compatible with thin-film technology. For a single metal strip of width $w$, length $l$, and thickness $t$, as shown in Fig.\,\ref{fig_inductance}(a), we have $r_{\rm GMD}\approx e^{-3/2}(w+t)$ and $r_{\rm AMD}=(w+t)/3$\,\cite{Grover1946, Greenhouse1974}. Thus, we express the self inductance as\,\cite{Greenhouse1974, Yue2000}
\begin{align}
	L_{\rm s} \approx \frac{\mu_{0}l}{2\pi}\left[\ln\left(\frac{2l}{w+t}\right) + \frac{1}{2} + \frac{w+t}{3l}\right].
\end{align}
For zero-thickness thin film ($t\rightarrow 0$), we have
\begin{align}
	L_{\rm s} \approx \frac{\mu_{0}l}{2\pi}\left[\ln\left(\frac{2l}{w}\right) + \frac{1}{2} + \frac{w}{3l}\right]. \label{eq:inductance_self}
\end{align}

Assuming that $l/w > 2.23$, the reciprocal term in the above equation is $10$ times smaller than the logarithmic term, and can be fairly neglected. Taking the partial derivative of $L$ with respect to $l$ and $w$, respectively,
\begin{align}
    \frac{\partial L_{\rm s}}{\partial l} &\approx 
    \frac{\mu_{0}}{2\pi}\left[\ln\left(\frac{2l}{w}\right) + \frac{3}{2}\right],\\
    \frac{\partial L_{\rm s}}{\partial w} &\approx -\frac{\mu_{0}}{2\pi}\frac{l}{w},
\end{align}
we find that $L_{\rm s}$ is more sensitive to the changes in $l$ than in $w$ for $l/w < 3.42$. 

\begin{figure}[t]
  \centering
  \includegraphics[width=\columnwidth]{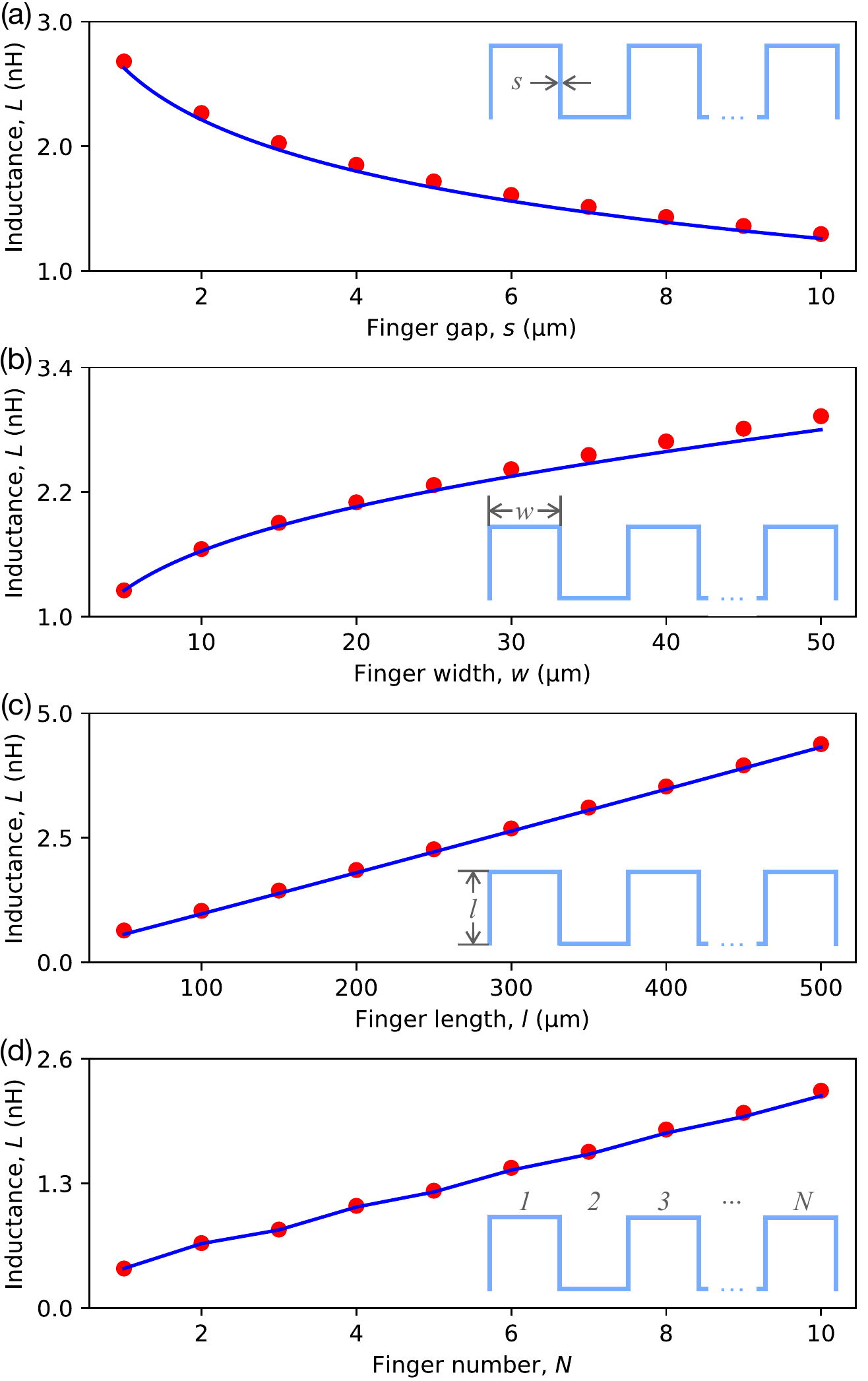}
  \caption{Inductance $L$ of a meander-line inductor (MLI) as a function of the (a) dielectric finger gap $s$, (b) width $w$, (c) length $l$, and (d) number $N$ for the derived analytical equations (blue curves) and the FEM simulations (red dots). The parameter values of the MLI are $s=2\,{\rm\mu m}$, $w=25\,{\rm\mu m}$, $l=250\,{\rm\mu m}$, and $N=10$ unless given in the panel. The insets show the MLI geometry with white denoting dielectrics and blue denoting the conductor.}
  \label{fig_inductor_FEM}
\end{figure}

\subsection{Mutual inductance}
In addition to self inductance, currents flowing in parallel lines lead to mutual inductance. Here, we consider two parallel wires of equal length, $l$, which are separated by a pitch distance, $s$, i.e., the distance between the centers of the two wires. The mutual inductance is given by \cite{Greenhouse1974, Yue2000}
\begin{align}
	L_{\rm m}(l, s) = \pm \frac{\mu_{0}l}{2\pi}Q(l, s),
	\label{eq:inductance_mutual_1}
\end{align}
where 
\begin{align}
	Q(l, s_{\rm GMD}) =& \ln\left[1
	+ \sqrt{1+\left(\frac{s_{\rm GMD}}{l}\right)^2}\right]
	- \ln\left(\frac{s_{\rm GMD}}{l}\right) \nonumber \\
	&+ \left[\frac{s_{\rm GMD}}{l}
	- \sqrt{1+\left(\frac{s_{\rm GMD}}{l}\right)^2}\right]
	\label{eq:inductance_mutual_2}.
\end{align}
Besides, $s_{\rm GMD}$ is the geometric mean distance of the two conductors, which can be well approximated as the pitch distance, $s_{\rm GMD} \approx s$, in normal cases\,\cite{Yue2000}. The positive (negative) sign is chosen if the current in the two lines are pointing in the same (opposite) direction \cite{Greenhouse1974}. We observe from the above equation that the absolute value of the mutual inductance, $|L_{\rm m}|$, increases when decreasing $s_{\rm GMD}$ owing to the increase of the flux linkage. Surprisingly, as pointed out by Ref.\,\cite{Yue2000}, the mutual inductance $L_{\rm m}$ does not depend on $w$. 

We note that the above results are valid for two identical parallel conductors with the same width $w$ and length $l$. For a more general case, where the two conductors have the same width, $w$, but different lengths, $l_1$ and $l_2$, we have
\begin{align}
	2L_{\rm m}(l_1, l_2,s) &= L_{\rm m}(l_2+l_3,s) 
	+ L_{\rm m}(l_1-l_3,s) \nonumber\\
	&- L_{\rm m}(l_3,s)
	- L_{\rm m}(l_1-l_2-l_3,s).
\end{align}
Here, we have assumed that $l_1 \geq l_2$ and the end of the second conductor is shifted by $l_3$ compared with the first one. Another situation is that the two conductors have the same length $l$ but different widths, $w_1$ and $w_2$. A careful calculation of the geometric mean distance, $s_{\rm GMD}$, indicates that\,\cite{Hsu2004}
\begin{align}
	\ln\left(s_{\rm GMD}\right) =
	& \ln(s) -\frac{3}{2}\nonumber\\
	&+ \frac{s^2}{2w_1 w_2}
	\left[\left(1+\frac{w_1+w_2}{2s}\right)^2\ln\left(1+\frac{w_1+w_2}{2s}\right)\right.\nonumber\\
	&\left. + \left(1-\frac{w_1+w_2}{2s}\right)^2\ln\left(1-\frac{w_1+w_2}{2s}\right)\right.\nonumber\\
	&\left. - \left(1+\frac{w_1-w_2}{2s}\right)^2\ln\left(1+\frac{w_1-w_2}{2s}\right)\right.\nonumber\\
	&\left. - \left(1-\frac{w_1-w_2}{2s}\right)^2\ln\left(1-\frac{w_1-w_2}{2s}\right)\right].
\end{align}
When $w_1, w_2 \ll s$, the value of $s_{\rm GMD}$ depends only weakly on the widths of the two strips. It is thus convenient to replace $s_{\rm GMD}$ by $s$ and neglect $w$ in general when calculating mutual inductance. 

\subsection{Numerical results}\label{sec:mli_fem}
To verify the applicability of the above equations in the typical parameter range of SQC, we numerically compare the analytical results with the FEM simulation for different control parameters. Here, we visualize a MLI as interdigital fingers made of the substrate material, as indicated in Fig.\,\ref{fig_inductor_FEM}, in order to keep the names of the variables consistent with the IDC that will be introduced in Sec.\,\ref{s:IDC}. In this way, the line width of the metal strip is defined as the finger gap, $s$, while the pitch distance is a summation of the finger width $w$ and gap $s$. Besides, $l$ defines the length of the finger and $N$ the number of fingers. The self and mutual inductances of the MLI can be estimated by combining Eqs.\,\eqref{eq:inductance_self}, \eqref{eq:inductance_mutual_1}, and \eqref{eq:inductance_mutual_2}. That is
\begin{align}
    L_{\rm s} &= NL_{\rm s}(w+s) + (N+1)L_{\rm s}(l),\\
    L_{\rm m} &= 2\sum_{n=1}^{N}\left(N+1-n\right)L_{\rm m}[l, n(w+s)].
\end{align}
The factor of two in the mutual inductance originates from the summation over the distance between two arbitrary parallel lines. The total mutual inductance has a component caused by the current flowing in each of the two lines. Note that the sign of $L_{\rm m}[l, n(w+s)]$ changes according to the relative orientations of the current, as indicated in Eq.\,\eqref{eq:inductance_mutual_1}. The total inductance is therefore readily obtained by adding the self and mutual inductances according to Eq.\,\eqref{eq:inductance}.

The FEM simulations are carried out in Keysight advanced design system (ADS). Our tests on other software such as Sonnet and Ansys HFSS also provide similar results. 
The substrate has a thickness of $675\,{\rm \mu m}$ and a relative permittivity of $\epsilon_{\rm r}=11.45$ (with a 250-nm ${\rm SiO_2}$ layer on top of Si). The superconducting layer is assumed to be a perfect conductor with vanishing thickness. Our numerical results are summarized in Fig.\,\ref{fig_inductor_FEM}. We observe an excellent agreement between the analytical equations and the FEM results for all the chosen $40$ sets of parameters. The observed consistency thus demonstrates the correctness of the modelling of the MLI, where the electromagnetic parameters are directly obtained from the layout.

We observe that the absolute value of the mutual inductance, $L_{\rm m}$, is generally two times smaller than the self inductance, $L_{\rm s}$, in all the above simulations. It therefore indicates that a spiral layout of the superconducting wire may enhance the total inductance, $L$, by a factor of around $0.5$, while a coplanar meandering-line layout suppress $L$ by $0.5$ compared with the pure self inductance. However, the major benefit of the meander-geometry is that it is easy to fabricate with the standard laser lithography tools in ordinary lab. Increasing the wire length may be the most effective way to improve the total inductance of the MLI.

\begin{figure}[t]
  \centering
  \includegraphics[width=\columnwidth]{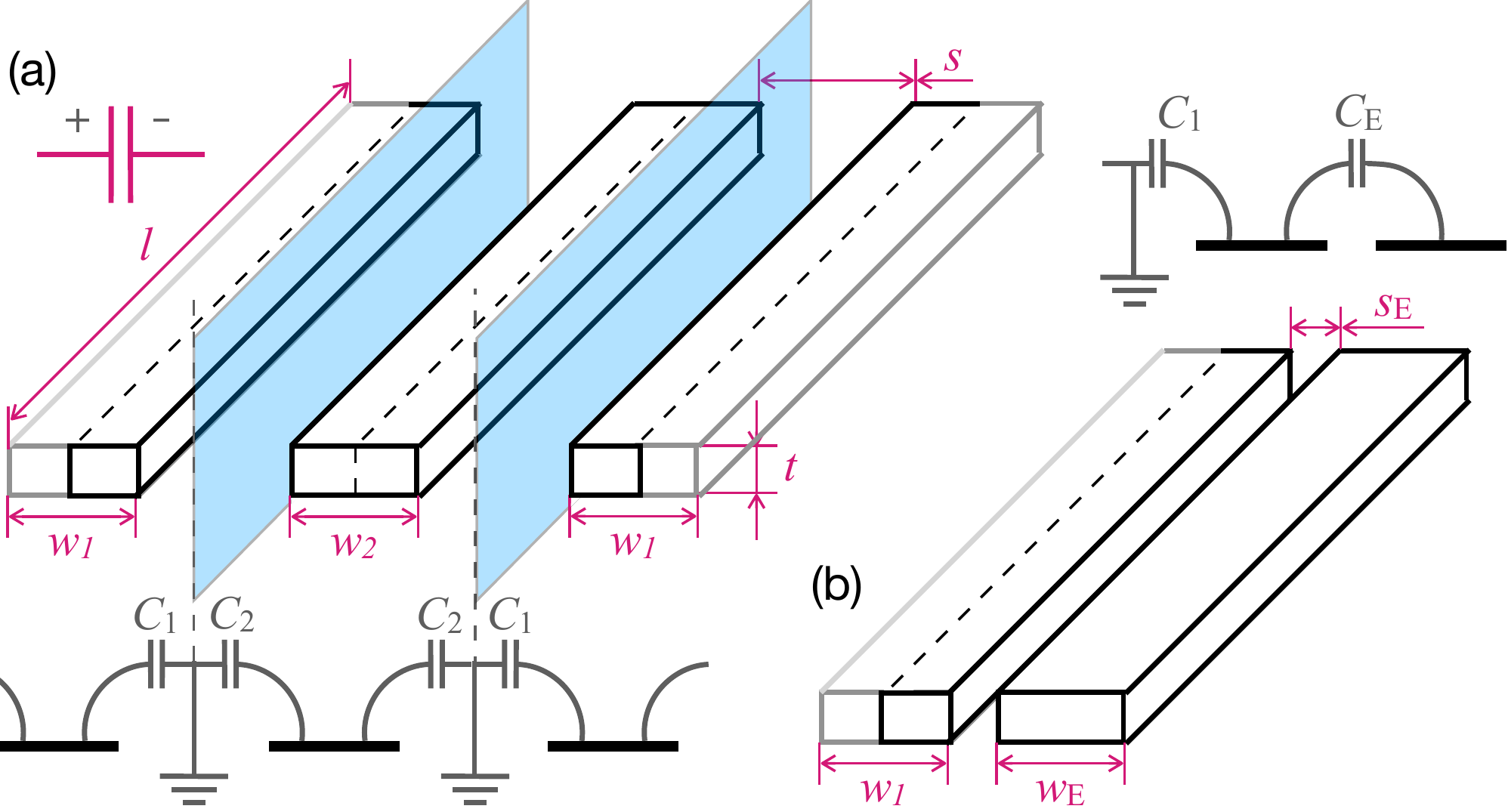}
  \caption{Illustration of capacitance for an interdigital capacitor (IDC). (a) The capacitance between two adjacent interior fingers with length $l$, thickness $t$, and widths $w_1$ and $w_2$, is calculated by summing over the capacitances, $C_1$ and $C_2$, with respect to the electrical wall with \textit{zero} potential in the center of the gap (gap size $s$). (b) The capacitance between the last interior finger and the exterior finger with width $w_{\rm E}$ is calculated directly without assuming an electrical wall in the middle of the gap (gap size $s_{\rm E}$).}
  \label{fig_capacitance}
\end{figure}

\begin{figure*}
  \centering
  \includegraphics[width=2\columnwidth]{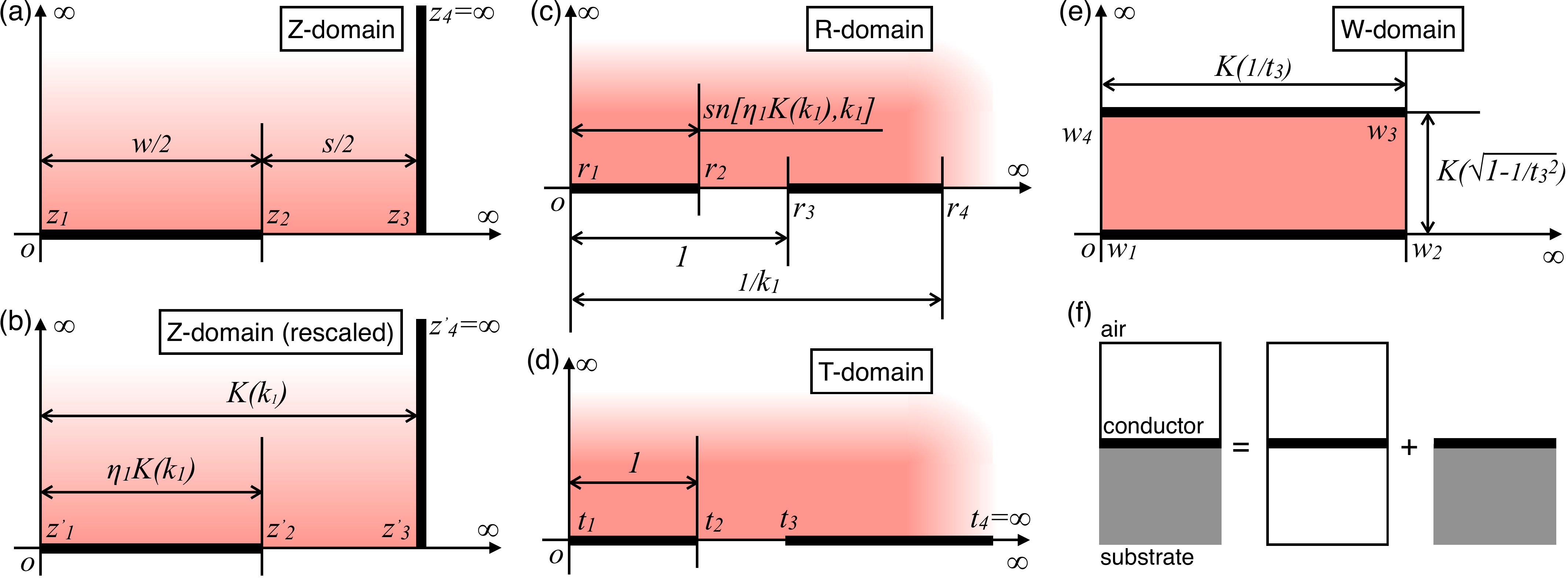}
  \caption{(a)--(e) Sequence of conformal mappings that transforms the original problem into a problem of calculating the capacitance between two parallel plates. See text in Sec.~\ref{sec:interior_fingers} for details. (f) Configuration for partial capacitances for a thin-film conductor sandwiched between air and the substrate.}
  \label{fig_mapping}
\end{figure*}

\section{Interdigital capacitor (IDC)}\label{s:IDC}
The IDC is made of multiple superconducting thin films which interdigitate with each other like interleaved fingers [Fig.\,\ref{fig_capacitance}(a)]. The layout is dual to MLI if one swaps the conductor and the dielectric material. Here, we define two types of interior fingers with different widths, $w_1$ and $w_2$, which are separated by a gap $s$. The width and the gap of the two exterior fingers are defined as $w_{\rm E}$ and $s_{\rm E}$, respectively, as shown in Fig.\,\ref{fig_capacitance}(b). They may be different from those of the interior fingers. All fingers have the same length $l$ and thickness $t$. We neglect the two pads that connect the fingers with the feedlines, since they are spatially far away from each other and play a negligible role in general\,\cite{Bao2019}.

To derive analytical relations between the circuit geometry and the capacitance, $C$, we assume that $(N-1)$ electrical walls with \textit{zero} potential are located in the middle of each slot which are orthogonal to the sample plane \cite{Wei1977}. Although this assumption can hardly be fulfilled in practical devices with a finite $N$, it is often valid when the slots are relatively narrow compared to the finger widths\,\cite{Igreja2004}. In this regard, we write the total capacitance of the IDC as
\begin{align}
	C = (N-3)\frac{C_1C_2}{C_1+C_2} + 2C_{\rm E}. \label{eq:capacitance}
\end{align}
Here, $C_1$ and $C_2$ are the capacitances between interior fingers, i.e., Type~1 and Type~2 width different widths, and the electrical wall. We define $C_{\rm E}$ as the capacitance between the interior finger (Type~1) and the exterior finger in each unit cell [Fig.\,\ref{fig_capacitance}(b)]. 

Below, we assume zero-thickness metal strips ($t=0$) to simplify the discussion. The finite-thickness case may be conveniently handled by resorting to Ref.\,\cite{Esfandiari1983}, where we apply the so-called Wheeler's first order approximation and redefine the finger width as 
\begin{align}
	w' = w + \frac{t}{\pi}\left[1+\ln\left(\frac{4\pi w}{t}\right)\right].
\end{align}

\begin{figure}[t]
  \centering
  \includegraphics[width=\columnwidth]{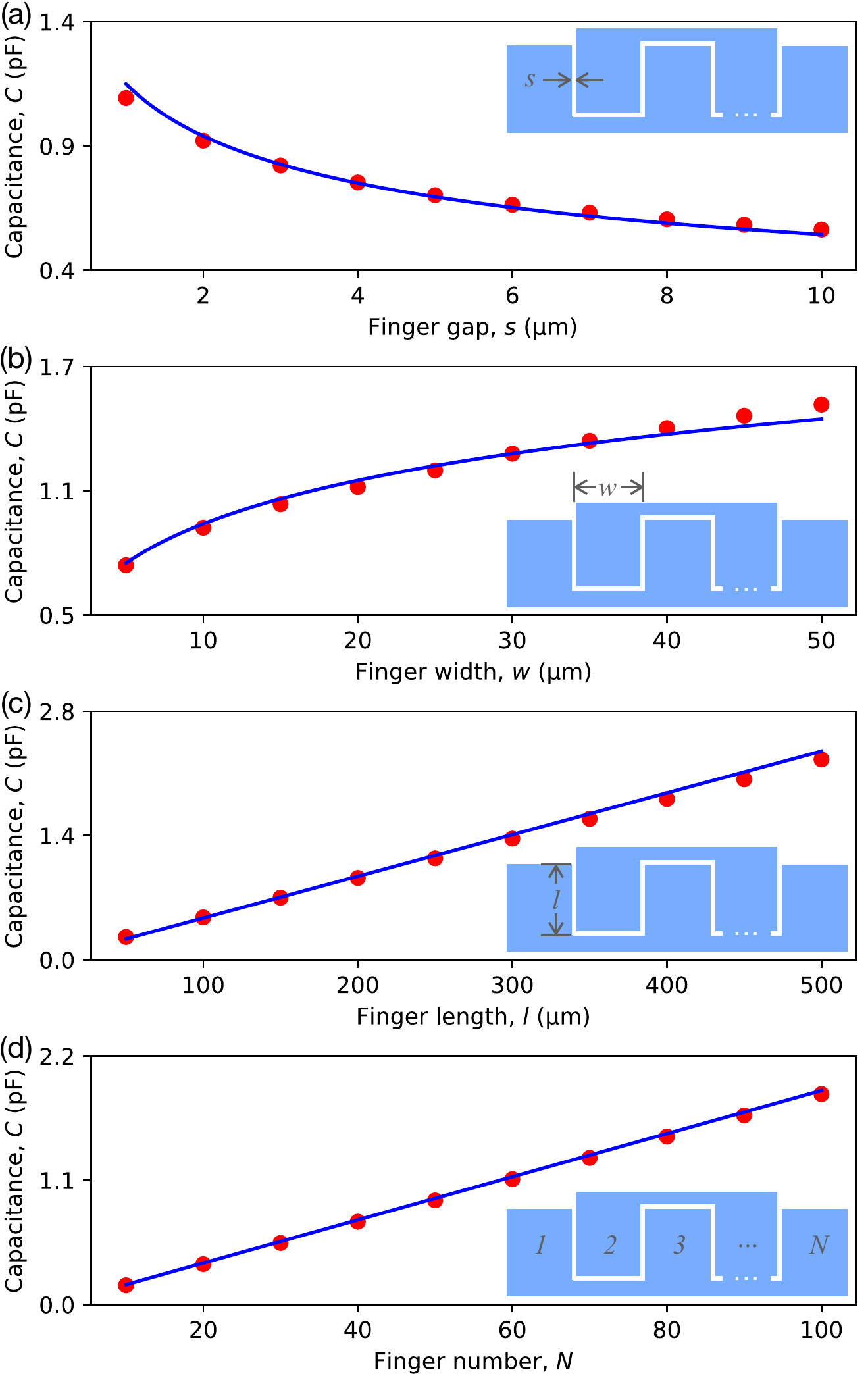}
  \caption{Capacitance $C$ of a interdigital capacitor (IDC) as a function of the (a) conductor finger gap $s$, (b) width $w$, (c) length $l$, and (d) number $N$ for the analytical equations of the main text (blue curves) and the FEM simulations (red dots). The parameter values of the IDC are $s=2\,{\rm\mu m}$, $w=10\,{\rm\mu m}$, $l=200\,{\rm\mu m}$, and $N=50$ unless given in the panel. The insets show the IDC geometry with white denoting dielectrics and blue denoting the conductor.}
  \label{fig_capacitor_FEM}
\end{figure}

\subsection{Interior fingers}\label{sec:interior_fingers}
To calculate the capacitance, e.g., $C_1$ and $C_2$, our major goal is to transform the coplanar geometry to an equivalent parallel plate geometry. The procedure is called conformal mapping, as illustrated in Fig.\,\ref{fig_mapping}(a)--(e). We start from the Z-domain where $z_1=0$, $z_2=w_1/2$, $z_3=w_1/2+s/2$, and $z_4=\infty$ [Fig.\,\ref{fig_mapping}(a)]. First, we rescale the quadrant in the $Z$-domain by a factor of $\alpha=K(k_1)/z_3$, such that $z'_1=0$, $z'_2=\eta_1 K(k_1)$, $z'_3=K(k_1)$, and $z'_4=\infty$ [Fig.\,\ref{fig_mapping}(b)], where we have named $\eta_1=w_1/(w_1+s)$ as the so-called metallization ratio\,\cite{Igreja2004}, $K(\cdot)$ is the the complete elliptic integral of the first kind, and $k_1$ is an arbitrary constant to be determined. It is convenient to define 
\begin{align}
	K(k_1)/K(k'_1) = (w_1+s)/(2h),
\end{align}
where $k'_1=\sqrt{1-k_1^2}$ and $h$ is the thickness of the substrate. It allows us to obtain a compact result when performing the following transformation from the Z-domain to the R-domain [Fig.\,\ref{fig_mapping}(c)]
\begin{align}
	r = {\rm sn}(z,k_1),
\end{align}
where ${\rm sn}(\cdot,k)$ is the Jacobi elliptic function of modulus $k$. This transformation gives $r_1=0$, $r_2={\rm sn}\left[\eta_1 K(k_1),k_1\right]$, $r_3=1$, and $r_4=1/k_1$.

Next, we apply the so-called M{\"o}bius transformation
\begin{align}
	t = \frac{r}{r_2}\sqrt{\frac{r_4^2-r_2^2}{r_4^2-r^2}},
\end{align}
which transforms the first quadrant in the $R$-domain to the first quadrant in the $T$-domain with $t_1=0$, $t_2=1$, and $t_4=\infty$ [Fig.\,\ref{fig_mapping}(d)]. Besides, we have
\begin{align}\label{eq:t3}
	t_3 = \frac{1}{{\rm sn}\left[\eta_1 K(k_1),k_1\right]}\sqrt{\frac{1-k_1^2{\rm sn}^2\left[\eta_1 K(k_1),k_1\right]}{1-k_1^2}}.
\end{align}

Finally, we apply the Schwarz--Christoffel mapping
\begin{align}
	w = \int_{0}^{t} \mathrm{d}t \frac{1}{\sqrt{(1-t^2)[1-(t/t_3)^2]}},
\end{align}
which folds the first quadrant of the T-domain into a rectangle in W-domain with width, $K$, and height, $K'$ [Fig.\,\ref{fig_mapping}(e)]. In particular \cite{Abramowitz1948}, 
\begin{align}
	K &= K(1/t_3) = \int_{0}^{1} \mathrm{d}t \frac{1}{\sqrt{(1-t^2)[1-(t/t_3)^2]}},\\
	K' &= K(\sqrt{1-1/t_3^2}) = \int_{1}^{t_3} \mathrm{d}t \frac{1}{\sqrt{(t^2-1)[1-(t/t_3)^2]}},
\end{align}
where $t_3$ is defined in Eq.\,\eqref{eq:t3}. For $h \gg w_1$, we have $k_1\rightarrow 0$ and $K(0)=\pi/2$ such that 
\begin{align}
	1/t_3 = \sin\left(\frac{\pi \eta_1}{2}\right). \label{eq:capacitance_t3}
\end{align}

By applying the described sequence of conformal transformations, we transform the original coplanar capacitor into a parallel plate capacitor with width $K(1/t_3)$, length $l$, and distance $K(\sqrt{1-1/t_3^2})$. The capacitance between the metal strip and the electrical wall through upper half plane in free space is thus 
\begin{align}
	C_{1,0} = \epsilon_0 l \frac{K(1/t_3)}{K(\sqrt{1-1/t_3^2})}, \label{eq:capacitance_C10}
\end{align}
with $\epsilon_{0} \approx 8.85\times 10^{-12}\,{\rm F/m}$ being the permittivity of free space. Similarly, the capacitance through the lower half-plane, which is filled with the substrate with dielectric constant $\epsilon_{\rm r}$, can be calculated as $C_{1,{\rm r}}=\epsilon_{\rm r}C_{1,0}$ [Fig.\,\ref{fig_mapping}(f)]. In summary, the total capacitance of the unit cell (Type~1) is
\begin{align}
	C_{1} = 2\epsilon_{\rm eff}C_{1,0}, \label{eq:capacitance_C1}
\end{align}
where $\epsilon_{\rm eff}=(1+\epsilon_{\rm r})/2$.

In the same way, one can calculate the total capacitance of the unit cell (Type~2) as $C_{2} = 2\epsilon_{\rm eff}C_{2,0}$. Here, we have
\begin{align}
	1/t_3 = {\rm sn}\left[\eta_1 K(k_2),k_2\right]\sqrt{\frac{1-k_2^2}{1-k_2^2{\rm sn}^2\left[\eta_2 K(k_2),k_2\right]}},
\end{align}
where $\eta_2=w_2/(w_2+s)$. It is approximately $1/t_3 = \sin\left(\pi \eta_2/2\right)$ for $h \gg w_2$. We note that we have neglected the edge effect when calculating the capacitance between two parallel plates. A more careful treatment requires the solution of the so-called Love equation, of which the analytical result with enough symmetry has been found only very recently\,\cite{Love1949, Reichert2020}.

\subsection{Exterior fingers}
One may follow the above procedure to derive $C_{\rm E}$ for the two cells with exterior fingers. Here, $C_{\rm E}=C_1C'_{\rm E}/(C_1+C'_{\rm E})$ with $C'_{\rm E} = 2\epsilon_{\rm eff}C'_{{\rm E},0}$, $C'_{\rm E} = \epsilon_0 l K(1/t_3)/K(\sqrt{1-1/t_3^2})$, and $1/t_3\approx \sin\left(\pi \eta_{\rm E}/2\right)$. The metallization ratio is defined as $\eta_{\rm E}=2w_{\rm E}/(2w_{\rm E}+s)$. Strictly speaking, this formula is valid only for $w_{\rm E} \simeq w_1, w_2$\,\cite{Igreja2004}. However, it is found to be rather precise in the practice of microwave engineering\,\cite{Bao2019}. 

For a more accurate result, we keep the last cell as a whole and define $z_1=0$, $z_2=w_1/2$, $z_3=w_1/2+s_{\rm E}$, $z_4=w_1/2+(s_{\rm E}+w_{\rm E})$, and $z_5=\textrm{i}h$ in the Z-domain. The transformation
\begin{align}
	r = \sinh \left(\frac{\pi z}{2h}\right)
\end{align}
maps the points to the R-domain, and results in $r_5=\textrm{i}$ and $r_1=0$\,\cite{Bao2019}. Following the transformations described above, we have
\begin{widetext}
\begin{align}
	1/t_3 &= \frac{\sinh[\pi w_1/(4h)]}{\sinh[\pi (w_1+2s_{\rm E})/(4h)]} 
	\sqrt{\frac{\sinh^2[\pi (w_1+2s_{\rm E}+2w_{\rm E})/(4h)]-\sinh^2[\pi (w_1+2s_{\rm E})/(4h)]}{\sinh^2[\pi (w_1+2s_{\rm E}+2w_{\rm E})/(4h)]-\sinh^2[\pi w_1/(4h)]}}.
\end{align}
\end{widetext}
Finally, for $h \gg w_1, w_{\rm E}$ we have 
\begin{align}
	1/t_3 = \frac{w_{1}/2}{w_1/2+s_{\rm E}}\sqrt{\frac{(w_1/2+s_{\rm E}+w_{\rm E})^2-(w_1/2+s_{\rm E})^2}{(w_1/2+s_{\rm E}+w_{\rm E})^2-(w_1/2)^2}}.
\end{align}
Correspondingly, we obtain $C_{\rm E} = 2\epsilon_{\rm eff}C_{{\rm E},0}$ and $C_{{\rm E},0} = \epsilon_0 l K(1/t_3)/K(\sqrt{1-1/t_3^2})$.

\subsection{Numerical results}
Similar to the study in MLI, we compare the analytical equations derived above with the numerical FEM results with different characteristic parameters, i.e., the finger number $N$, gap width $s$, and finger length $l$, and finger width $w$. The total capacitance can be estimated by combining Eqs.\,\eqref{eq:capacitance}, \eqref{eq:capacitance_t3}, \eqref{eq:capacitance_C10}, and \eqref{eq:capacitance_C1}, where the metallization ratios for the three different unit cells are specified individually. The parameters $N$, $s$, $l$, and $w$ are varied in ranges $[10,100]$, $[1\,{\rm \mu m},10\,{\rm \mu m}]$, $[50\,{\rm \mu m},500\,{\rm \mu m}]$, and $[5\,{\rm \mu m},50\,{\rm \mu m}]$. The comparison results are summarized in Fig.\,\ref{fig_capacitor_FEM}. The other parameters of the FEM solver are set as for the MLI in Sec.\,\ref{sec:mli_fem}. The excellent agreement between the analytical and the FEM results for all the chosen 40 geometries indicates the correctness of our modelling in this parameter range relevant for SQC. 

\section{Compact LC resonators}
With the above-derived knowledge of the MLI and the IDC, we connect them in parallel to design compact LC resonators without FEM. Here, we are particularly interested in the sub-gigahertz regime since it is not conveniently reachable by compact CPW resonators. The sub-gigahertz resonators also have an immediate application in superconducting thermal detectors, such as bolometers and calorimeters\,\cite{Govenius2016, Kokkoniemi2019, Kokkoniemi2020, Gasparinetti2015, Karimi2020}.  

\begin{figure}[h!]
 \centering
 \includegraphics[width=\columnwidth]{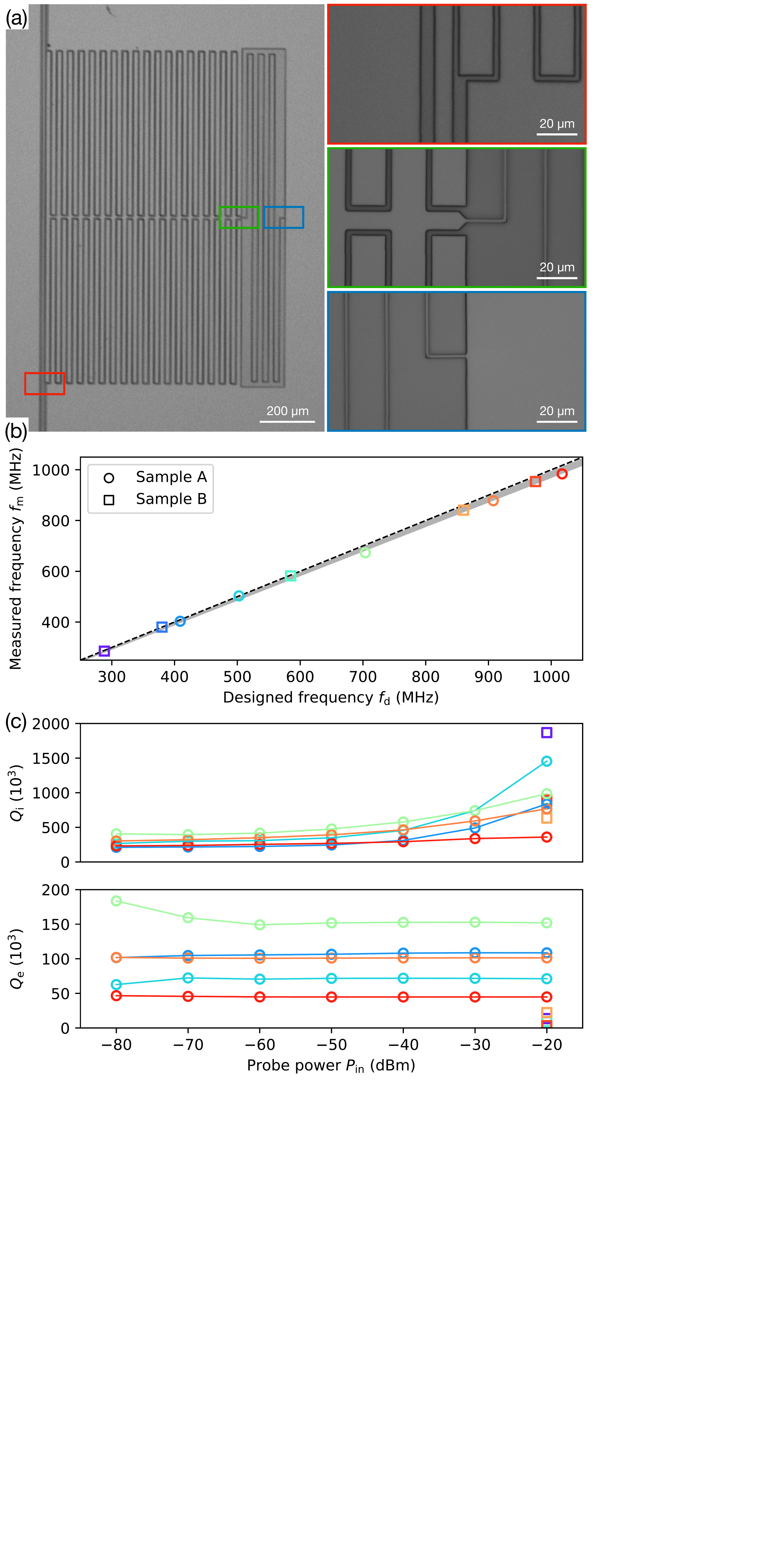}
 \caption{Characterization results of the lumped LC resonators. (a) Optical photograph of one resonator in Sample B with designed resonance frequency $\omega_{0}/2\pi=600\,{\rm MHz}$. The insets show the enlarged area with better resolution. The resonators in Sample A have a $\sim 5$-times lower coupling strength to the waveguide by design. (b) Comparison between the designed and measured resonance frequencies, $f_{\rm d}$ and $f_{\rm m}$, respectively. The dashed line indicates the perfect case where $f_{\rm d}=f_{\rm m}$, while the shaded area indicates the standard deviation around the achieved mean-value frequency. (c) Extracted values of the internal and external Q factors, $Q_{\rm i}$ and $Q_{\rm e}$, under different probe power $P_{\rm in}$. Here, the colors denote different resonators.}
 \label{fig_LC_experiment}
\end{figure}

\subsection{Experimental methods}
We fabricate $10$ compact LC resonators on two pure Si chips with an area of $10\times 10\,{\rm mm^2}$, denoted as Sample A and B as showcased in Fig.\,\ref{fig_LC_experiment}(a). The sample substrate has a thickness of $675\,{\rm \mu m}$. A $200$-${\rm nm}$-thick Nb layer is applied by sputtering, and then patterned by maskless laser lithography and reactive ion etching (RIE). The building block for the resonators in Sample A are the MLI and IDC with the same parameter $s=3\,{\rm \mu m}$, $w=22\,{\rm \mu m}$, and $l=1300\,{\rm \mu m}$. For Sample B, we have $s=2\,{\rm \mu m}$, $w=18\,{\rm \mu m}$, and $l=1200\,{\rm \mu m}$. In both cases, we adjust the number of these modules to reach the desired resonance frequency and impedance as close as possible. The relative permittivity of the substrate is chosen as $\epsilon_{\rm r}=11.9$ for design. The resonators are coupled to a common transmission line for multiplexed readout, such that each resonator forms a hanger-type geometry \cite{Chen2022, Chen2022b}. On each chip, the designed resonance frequencies increase from $300\,{\rm MHz}$ to $1\,{\rm GHz}$ separated by approximately $100\,{\rm MHz}$. The designed impedances of the four low-frequency resonators increase from $30\,{\Omega}$ to $60\,{\Omega}$ with the frequency, while they are kept below $30\,{\Omega}$ for the four high-frequency resonators. However, we observe only $5$ resonance peaks in each sample which we attribute to the possibly uneven RIE in the fabrication process.

We thermalize the samples at approximately $30\,{\rm mK}$ for characterization. The input line has approximately $60\,{\rm dB}$ of attenuation from room to cryogenic temperatures. We extract the resonance frequency and the quality factors of each individual resonator from the transmission coefficients. In the vicinity of a single resonance frequency, we describe the measured transmission coefficient as\,\cite{Chen2022, Chen2022b}
\begin{align}
    S_{21}(\omega) = A\textrm{e}^{-\textrm{i}(\omega \tau + \varphi)}\left(1 - \frac{\textrm{e}^{\textrm{i}\phi}Q_{\rm l}/Q_{\rm e}}{1 - \textrm{i}2Q_{\rm l}\delta}\right),
\end{align}
where $\delta=(\omega-\omega_{\rm r})/\omega_{\rm r}$, and $\omega_{\rm r}$ is the resonance frequency that may be slightly different from the designed bare resonator frequency $\omega_{0} = 1/\sqrt{LC}$ because of the coupling. The loaded, internal, and external quality factors are defined as $Q_{\rm l}$, $Q_{\rm e}$, and $Q_{\rm i}$, respectively, which satisfy $1/Q_{\rm l} = 1/Q_{\rm i} + 1/Q_{\rm e}$. The constant parameters $A$, $\tau$, $\varphi$, and $\phi$ are related to the practical distortions to the measured spectrum. After correcting these distortions, the retrieved complex-valued spectrum forms a circle which is centered on the real axis and passes through a fixed point $(1+\textrm{i}0)$ for $\omega \rightarrow \infty$. The radius of the circle is $Q_{\rm l}/(2Q_{\rm e})$, while $Q_{\rm l}$ can be conveniently obtained by fitting the lineshape with a Lorentzian function. 

\subsection{Experimental results and analysis}
Figures\,\ref{fig_LC_experiment}(b)--(c) summarize the measurement results. The excellent agreement between the designed and the extracted resonance frequencies indicates the accurate control of the resonance frequency [Fig.\,\ref{fig_LC_experiment}(b)], and therefore demonstrates the high accuracy of our analytical model consisting of the lumped-element components. On average over the $10$ measured resonators, the experimentally obtained resonance frequency is $1.83\%$ below the design frequency with a standard deviation of $1.44\%$. This small difference may have several possible origins such as the coupling between the resonator and the waveguide, which unavoidably shifts the actual resonance frequency from the design value $\omega_{0}/2\pi$ to a lower value. On the other hand, the shift can also arise from possible fabrication imperfections such as over etching of the Nb thin film during RIE, the possible under estimation of the dielectric constant at low temperature, and the small contribution of the kinetic inductance that has been neglected in the resonator design. We also note that, as indicated in the FEM simulations shown in Figs.\,\ref{fig_inductor_FEM} and \ref{fig_capacitor_FEM}, the introduced analytical equations tend to underestimate especially the total inductance. 

The extracted quality factors indicate a relatively long coherence time of the compact LC resonators [Fig.\,\ref{fig_LC_experiment}(c)]. At high probe power ($P_{\rm in}=-20\,{\rm dBm}$ at the room temperature), the average internal quality factor of all the $10$ resonators is $944\times 10^{3}$ with a standard deviation $405\times 10^{3}$. These values decrease with decreasing probe power but saturate at roughly $-60\,{\rm dBm}$. At the lowest probe power ($P_{\rm in}=-80\,{\rm dBm}$), we obtain $283\times 10^{3}$ and $69\times 10^{3}$ for the mean internal quality factor and its standard deviation, respectively. We note that high-quality CPW resonators, in the $4$--$8\,{\rm GHz}$ range with $Q_{\rm i} \geq 1000\times 10^{3}$ at low probe power, can be routinely implemented. However, our achieved value of $Q_{\rm i}$ in the sub-gigahertz range has been seldom reached by CPW resonators. Moreover, our samples are fabricated in a single step without sophisticated techniques that are normally required for fabricating a high-Q CPW resonator, such as the wet-chemical post-processing approach for creating a clean and smooth interface between the thin film and air \cite{Martinis2005, Gao2008, Romanenko2017, Mueller2019, McRae2020}.

\section{Conclusions}
We provided a systematic study of meander-line inductors (MLIs) and interdigital capacitors (IDCs) in the context of superconducting quantum circuits (SQC), and combined them to design compact LC resonators at sub-gigahertz frequencies. The experimental results show an excellent agreement between the FEM simulations and the simple analytical equations provided in this study, where the resonance frequency was experimentally demonstrated with only about $2\%$ deviation from the analytically obtained design value. These results promote the view that lumped-element inductors and capacitors can be accurately designed without resource-demanding FEM simulations. Consequently, these lumped-element resonators can be readily added to the toolkit of the SQC technology.

The physical size of a compact LC resonator is much smaller than the wavelength, which is in strong contrast to the CPW resonators. The usefulness of this compactness is pronounced at low frequencies where both the longitudinal and transverse dimensions can be tuned in a relatively large range below the wavelength. The feasibility of such a compact design is supported by the concept of photonic crystals, also referred to as metamaterials, where sub-wavelength structures are arranged periodically to tailor the behavior of the electromagnetic waves in a way that is beyond the capabilities of the uniform base material\,\cite{Joannopoulos1997, Joannopoulos2008, Skorobogatiy2009}. In our case, the unit cells are the $1300$-${\rm \mu m}$-high $25$-${\rm \mu m}$-wide patterns in the MLIs and IDCs.

The achieved compactness has a direct use in the current SQC technology. On one hand, fitting more circuit components on a single chip is a practical requirement when building large-scale quantum systems such as quantum computers or microwave photon lattices for quantum simulations\,\cite{Ozawa2019}. The shown sub-gigahertz resonators already find important applications in cryogenic particle detection\,\cite{Govenius2016, Kokkoniemi2019, Kokkoniemi2020, Gasparinetti2015, Karimi2020}. On the other hand, the sub-wavelength physical size confines the microwave photons in a small area which avoids the technical problem of keeping a precise impedance over an entire CPW resonator. The small mode volume naturally leads to a high and readily engineerable coupling strength to a nearby device, such as a qubit\,\cite{McKay2015}. The single-mode nature of lumped-element circuits also avoids the spurious effects of higher harmonics to the qubit lifetime and coherence, which is a known issue in SQC\,\cite{Houck2008}. We may therefore expect a change of paradigm in circuit design for large-scale quantum information processing by combing the capabilities of compact LC resonators and superconducting qubits.  

\section*{Acknowledgements}
This work is supported by  the Academy of Finland Centre of Excellence program (No.\,$336810$), European Research Council under Advanced Grant ConceptQ (No.\,$101053801$),  Business Finland Foundation through Quantum Technologies Industrial (QuTI) project (No.\,41419/31/2020), Technology Industries of Finland Centennial Foundation, Jane and Aatos Erkko Foundation through Future Makers program and Finnish Foundation for Technology Promotion (No.\,$8640$). We thank Leif Gr{\"o}nberg for niobium deposition.

\bibliography{LC_REF}  
\end{document}